\documentclass[%
reprint,
superscriptaddress,
 amsmath,amssymb,
 aps,
pre,
floatfix,
]{revtex4-2}

\usepackage{dcolumn}
\usepackage{bm}
\usepackage{comment} 
\usepackage{hyperref}
\usepackage{amssymb,amsmath,graphicx,array,color,amsthm}
\usepackage{caption}
\usepackage{amscd}
\usepackage[normalem]{ulem}
\usepackage{nccmath}
\usepackage{float}
\usepackage{array, multirow}
\usepackage{tabularx}
\usepackage{hhline}
\def \nradk {n^{\text{rad}}(k)}
\def \nradkt {n^{\text{rad}}(k,t)}
\def \nrad {n^{\text{rad}}}
\def \rmd {\,\mathrm{d}\,}
\def \kcf {k_{\rm cf}}
\newcommand{\etal}{\textit et al.~}

\begin{document}
\preprint{AIP/123-QED}

\title{Self-similar evolution of wave turbulence in Gross-Pitaevskii system}

\author{Ying Zhu}
\email{yzhu@unice.fr}
\affiliation{Universit\'{e} C\^{o}te d'Azur, CNRS, Institut de Physique de Nice (INPHYNI), 17 rue Julien Laupr\^{e}tre 06200 Nice, France}
\author{Boris Semisalov}
\affiliation{Universit\'{e} C\^{o}te d'Azur, Observatoire de la C\^{o}te d'Azur, CNRS, Laboratoire Lagrange, Boulevard de l'Observatoire CS 34229 -- F 06304 Nice Cedex 4, France}
\affiliation{Sobolev Institute of Mathematics SB RAS, 4 Academician Koptyug Avenue, 630090 Novosibirsk, Russia}
\affiliation{Federal Research Center for Information and Computational Technologies, 6 Academician Lavrentyev Avenue, 630090 Novosibirsk, Russia}
\author{Giorgio Krstulovic}
\affiliation{Universit\'{e} C\^{o}te d'Azur, Observatoire de la C\^{o}te d'Azur, CNRS, Laboratoire Lagrange, Boulevard de l'Observatoire CS 34229 -- F 06304 Nice Cedex 4, France}
\author{Sergey Nazarenko}
\affiliation{Universit\'{e} C\^{o}te d'Azur, CNRS, Institut de Physique de Nice (INPHYNI), 17 rue Julien Laupr\^{e}tre 06200 Nice, France}

\begin{abstract}
We study the universal non-stationary evolution of wave turbulence (WT) in Bose-Einstein condensates (BECs).
Their temporal evolution can exhibit different kinds of self-similar behavior corresponding to a large-time asymptotic
of the system or to a finite-time blowup.
We identify self-similar regimes in BECs by numerically simulating the forced and unforced Gross-Pitaevskii equation (GPE) and the associated wave kinetic equation (WKE) for the direct and inverse cascades, respectively. 
In both the GPE and the WKE simulations for the direct cascade, we observe the first-kind self-similarity that is fully determined by energy conservation. For the inverse cascade evolution, we verify the  existence of a self-similar
evolution of the second kind describing self-accelerating dynamics of the spectrum leading to blowup at the zero mode (condensate) at a finite time. 
We believe that the universal self-similar spectra found in the present paper are as important and relevant for understanding the BEC turbulence in past and future experiments as the commonly studied stationary Kolmogorov-Zakharov (KZ) spectra.
\end{abstract} 

\maketitle  

\section{Introduction}
A great number of physical wave systems exhibit states with broadband spectra of excited mutually interacting modes. Such states are called
Wave Turbulence (WT)~\cite{ZLF,nazarenko2011wave,galtier_2022}, and their examples can be found in classical fluids \cite{Zakharov:1967aa,Caillol_KineticEquationsStationary_2000,Galtier_WeakInertialwaveTurbulence_2003}, quantum and optical media \cite{dyachenko1992optical,Lvov_WeakTurbulenceKelvin_2010,proment2012sustained} and even in primordial Universe~\cite{galtier2017turbulence}. 
Reference to turbulence in WT occurs because, like in the case of classical hydrodynamics, the WT systems are typically characterized by 
self-similar cascades of the energy (or another invariant) through scales.
Associated with these cascades, there exist stationary self-similar spectra, the so-called Kolmogorov-Zakharov (KZ) spectra, that are analogous to the famous Kolmogorov spectrum of hydrodynamic turbulence, and that are expected in the forced-dissipated WT systems. Search and validation of the KZ spectra, theoretically, numerically and experimentally, has dominated most of the works on WT \cite{proment2012sustained,Galtier_WeakInertialwaveTurbulence_2003,Caillol_KineticEquationsStationary_2000,Lvov_WeakTurbulenceKelvin_2010,navon2019synthetic,Zakharov:1967aa,NLanrev,Griffin2022Energy,zhu2022direct}.  
This interest is explained by the universality of the KZ spectra, i.e. their insensitivity 
to  fine details of forcing and dissipation mechanisms (similar to the universality of the classical Kolmogorov spectrum).

On the other hand, temporal evolution leading to the formation of the KZ spectra, 
as well as the spectrum evolution in unforced systems, can also be universal and
exhibit self-similarity. Moreover, such non-stationary solutions are often more relevant in WT realized in laboratories and in natural situations. Self-similar behavior is rather nontrivial and comes in different types corresponding to an infinite-time asymptotic of the system or to a finite-time blowup. Moreover, the same wave system may simultaneously exhibit different kinds of self-similarity in different scale ranges. 

In the present paper, we report on a systematical study of non-stationary solutions arising in the forced and unforced Gross-Pitaevskii equation and the associated wave-kinetic equation furnished by the WT theory. In each of the considered settings, we keep our focus on identifying self-similar evolution regimes. Namely, we consider the following representative settings: forced-dissipated direct and inverse cascades and unforced-undissipated (free-decaying) direct and inverse cascades. In the forced-dissipated direct cascade, the energy is injected at low and dissipated at large wave numbers, and in the forced-dissipated inverse cascade, the particles are injected at large and dissipated at low wave numbers. In the unforced-undissipated systems, the direct and inverse cascades arise during a conservative (energy and particle preserving) evolution of spectra toward the high- and low-frequency ranges, respectively.
Note that in the unforced-undissipated settings, the solution of the wave-kinetic equation blows up in a finite time $t^*$ marking a non-equilibrium onset of the Bose-Einstein condensation (BEC) into the zero wave number mode \cite{semikoz1995kinetics}. Respectively, the WT description fails close to $t^*$ in the low-frequency range due to an accelerated nonlinear (and decelerated linear) dynamics, but the Gross-Pitaevskii evolution continues beyond this time without a blowup. In the forced-dissipated settings, the wave-kinetic blowup can be prevented by introducing dissipation of the zero- and low-frequency modes. In this case, the subsequent evolution leads to the formation of the stationary KZ spectrum \cite{zhu2022direct}. In this paper, we will aim at systematizing the previous and new findings about the non-stationary evolution of the BEC WT and at presenting a classification of the typical scenarios. Further, we will discuss our results from the point of view of novel perspective designs of the BEC turbulence experiments.

\section{Theoretical background}

\subsection{Gross-Pitaevskii model}
Gross-Pitaevskii equation (GPE) describes the evolution of ultra-cold Bosonic gases with repelling interaction potential \cite{pitaevskii2003bose}.
For our study, it suffices to work with the dimensionless GPE for the complex wave function $\psi({\bf x},t)$:
 \begin{equation}
   \frac{\partial\psi ({\bf x},t)  }{\partial t}=i \left [ \nabla^{2}  -\left|\psi({\bf x},t) \right|^{2}  \right ]\psi({\bf x},t) \,.
  \label{eq:GPE}
 \end{equation} 
We shall study numerically quasi-homogeneous quasi-isotropic turbulence   of weakly-interacting three-dimensional BEC in a triply-periodic cube of side $L$ (and of the volume $V=L^3$). 
The GPE \eqref{eq:GPE} conserves the total number of particles and energy per unit volume,
\begin{subequations}
\begin{equation}\label{eq:cons-N}
N =\frac{1}{V}\int_V |\psi(\mathbf{x}, t)|^2 \rmd\mathbf{x}\,, 
\end{equation}
\begin{equation}\label{eq:cons-H}
 H=\frac{1}{V}\int_V \left[ |\nabla \psi(\mathbf{x}, t)|^2 + \frac{1}{2}|\psi(\mathbf{x}, t)|^4 \right] \rmd\mathbf{x} \,, 
\end{equation}
\end{subequations}
respectively.

\subsection{Wave Turbulence theory}
When the zero-frequency mode (uniform condensate) is negligible, the WT theory for the GPE formulates an asymptotic closure for the waveaction spectrum $n_{\bf k}(t)\equiv n({\bf k},t) = \frac{V}{(2\pi)^3}\langle |\hat \psi_{\bf k}(t)|^2 \rangle,$
where $\hat \psi_{\bf k}(t)$ is the Fourier transform of
$\psi(\mathbf{x}, t)$, and the brackets denote averaging over the initial wave statistics.
The WT closure is derived under assumptions of small nonlinearity and random initial phases and amplitudes of waves~\cite{ZLF,nazarenko2011wave}. It furnishes a wave-kinetic equation (WKE) with four-wave interactions \cite{ZAKHAROV1985285,dyachenko1992optical}. For an isotropic spectrum, which depends only on the magnitude of the wave vector $k=|{\bf k}|$, it is given by
\begin{equation}
\begin{split}\label{eq:WKE}
&\frac{\partial }{\partial t} n_{\omega}(t)= \frac{4\pi^3}{\sqrt{\omega}}\int \min\left(\sqrt{\omega},\sqrt{\omega_1},\sqrt{\omega_2},\sqrt{\omega_3}\right)
\delta(\omega^{01}_{23}) \\
&\times n_{\omega}  n_{1}n_{2}n_{3} 
\left(n_{\omega}^{-1} + n_{1}^{-1} - n_{2}^{-1} - n_{3}^{-1}\right)
\mathrm{d}\omega_1\mathrm{d}\omega_2\mathrm{d}\omega_3,
\end{split}
\end{equation}
where $\omega$ is the wave frequency determined by the dispersion relation $\omega =k^2$, 
 $n_{\omega}(t)=n(\omega,t)=n_{\bf k}(t)=n_k(t)$,
 $\omega^{01}_{23} \equiv \omega+\omega_1-\omega_2-\omega_3$, $\delta$ is the Dirac delta function. The integral in \eqref{eq:WKE} is taken over $\omega_{1},\omega_{2},\omega_{3}>0$.

 In this paper, we focus on the spherically-integrated wave-action spectrum $\nradk=4\pi k^2 n_{\omega}$, which is the spectral particle density depending on the wave vector radius $k$.
 The WKE for $\nrad_k=\nradk=\nradkt$ reads
  \begin{equation}\label{eq:WKE1}
  \begin{split}
    &\frac{\partial \nrad_k}{\partial t}= 2\pi\int
    \frac{\min\left(k,k_1,k_2,k_3\right)}{k\,k_1k_2k_3}
    n_{k}^{\rm rad} n_{{k_1}}^{\rm rad}n_{{k_2}}^{\rm rad}n_{{k_3}}^{\rm rad}  \\
    & \times \delta(\omega^{01}_{23})
    \left(\tfrac{k^2}{n_{k}^{\rm rad}} + \tfrac{k_1^2}{n_{{k_1}}^{\rm rad}} - \tfrac{k_2^2}{n_{{k_2}}^{\rm rad}}-\tfrac{k_3^2}{n_{{k_3}}^{\rm rad}}\right)  \mathrm{d}k_1\mathrm{d}k_2\mathrm{d}k_3\,. 
    \end{split}
    \end{equation}  
WKE conserves the total number of particles and the energy,
\begin{subequations}\label{eq:cons-WKE}
\begin{equation}
N = \int_0^\infty \nradk \, \mathrm{d}k \,, \label{eq:cons-Nwke}
\end{equation}
\begin{equation}
E = \int_0^\infty  k^2 \, \nradk \,\mathrm{d}k\,, \label{eq:cons-Ewke}
\end{equation}
\end{subequations}
which coincide with \eqref{eq:cons-N} and with the first term in \eqref{eq:cons-H} (the second term is small in WT), respectively.

Most central in the WT studies have been the KZ spectra: stationary solutions of the WKE, each realizing a constant spectral flux of an invariant of the system. For the GPE equation, there are two such invariants, $N$ and $E$ and, respectively, there are two KZ spectra. These spectra were proposed in \cite{dyachenko1992optical} and discussed in many papers since \cite{proment2009energy,proment2012sustained,navon2016emergence,navon2019synthetic}
but a rigorous systematic derivation of them, including finding the dimensionless pre-factors, was only done recently in \cite{zhu2022direct}.
In that work we obtained and validated numerically, using alongside the GPE and WKE simulations, the following KZ spectra corresponding to the direct energy and the inverse particle cascades ---
 \begin{equation} \label{eq:dc}
 \begin{split}
 {\rm direct:} \quad &\nradk =4\pi C_{\rm d} P_0^{1/3}  k^{-1} \ln^{-1/3}\left({k}/{k_{\rm f}}\right)\,,\\
 &C_{\rm d} \approx 5.26\times10^{-2}\,, 
 \end{split}
\end{equation}
\begin{equation}  \label{eq:ic}
\begin{split}
{\rm inverse:} \quad &\nradk =4\pi C_{\rm i} |Q_0|^{1/3} k^{-1/3}\,,\\
&C_{\rm i} \approx 7.5774045\times 10^{-2}\,,
 \end{split}
\end{equation}
 where  $P_0$ and $Q_0$ are the respective (constant) spectral fluxes of energy and particles through the sphere of radius $k=|{\bf k}|$. Note that the log-factor in spectrum \eqref{eq:dc} is due to the fact that the pure power law $\nradk\sim k^{-1}$ ($n_{\omega}\sim \omega^{-3/2}$) resulting from a dimensional argument corresponds to a marginally divergent integral in the WKE. As a remedy for this situation, the log-correction was suggested on phenomenological grounds in \cite{dyachenko1992optical} and proved rigorously in \cite{zhu2022direct}. 
 
 The direct and the inverse cascade spectra are different from each other in the following sense. 
 Assuming that the inertial range tends to infinity in the direct cascade spectrum, i.e. that this spectrum
 extends from the forcing wave number to arbitrarily high $k$'s, the energy integral \eqref{eq:cons-Ewke}
is divergent at the upper limit, $k \to \infty$. This corresponds to an infinite energy in physical space,
and the respective KZ spectrum is said to have an infinite capacity. 
On the other hand, for the inverse cascade spectrum with an infinite (in the $\log k$ variable) inertial range extending to $k\to0$ (hence $\log k \to -\infty$), the $N$-integral in \eqref{eq:cons-Nwke}
is finite (convergent at  $k\to0$). Such a spectrum is said to have a finite capacity, and it corresponds to a finite particle density in the physical space.

Finally, note that because of the $\delta(\omega^{01}_{23})$ term in the WKE \eqref{eq:WKE1},  if $n_{k}^{\rm rad}=$~const then the integral exactly vanishes. This equilibrium (no-flux) spectrum corresponds to thermodynamic energy equipartition and it is known as the Rayleigh–Jeans spectrum. It represents a special case of more general thermodynamic equilibrium:
\begin{equation}\label{eq:thermal}
\text{Rayleigh–Jeans:} \quad \nradk= \frac{4\pi k^2 T }{k^2+\mu},
\end{equation}
where $T$ and $\mu$ are two (positive) Lagrange multipliers that fix the total energy and mass. They can be interpreted as temperature and chemical potential and play an important role to understand the process of condensation \cite{Connaughton2005,KrstulovicDispersiveBottleneck,KrstulovicTruncatedGP}. Note that, in order to make the energy and mass finite, one needs to impose a UV-cutoff $k_{max}$ in the system. We will come back to this point later.

\section{Self-similar solutions} \label{Sec:3}

In Zeldovich's classification, the self-similar solutions can be of two kinds.
For the first kind of self-similarity, the self-similar coefficients are fully determined by a conservation law, e.g. energy --- like in the spherical shock wave resulting from a point-like energy deposition in an ideal gas. 
For the second kind of self-similarity, the self-similar coefficients cannot be determined by a conservation law only because most of the respective invariant remains in a volume that is not self-similar --- like in the problem of a spherical implosion of a vacuum bubble in gas. Yet, there is also a third kind of self-similarity in which the self-similar coefficients are fixed by a previous evolution stage which is also self-similar \cite{Bell_2018,Nazarenko2019}. 

In what follows we will see that all three kinds of self-similarity are relevant
to the evolving BEC WT. As a general guideline, one should expect the first kind
when the respective KZ spectrum has an infinite capacity and the second kind of self-similarity in the finite capacity case. Since in the BEC WT case the direct and inverse cascades have an infinite and finite capacity respectively, they  exhibit the first and the second kind self-similarities, respectively. 
This means, in particular, that the inverse cascade front reaches zero frequency in finite time $t^*$ setting a power law with an anomalous  (different from KZ) exponent. This is followed by a reflected-wave spectrum at $t>t^*$ propagating the KZ exponent toward the larger frequencies if dissipation is present at low frequencies. If there is no low-frequency dissipation, for large values of $t$ one gets the thermodynamic energy equipartition exponent for GPE. In both cases, the reflected wave is described by a self-similar solution of the third kind.

\subsection{The first-kind self-similarity for the direct cascade}

Assume that the self-similar solution of \eqref{eq:WKE1} has the following form, 
\begin{equation}\label{eq:self_dirc}
\nrad(k,t)=t^{-a}f(\eta)\quad \text{with}\quad \eta=k/t^b\,,
\end{equation}
Substituting the above expression into WKE (\ref{eq:WKE1}), we get
\begin{multline}
-(a f(\eta)+b\eta f'(\eta))t^{2a-1}= 2\pi\int  
\frac{\min(\eta,\eta_1,\eta_2,\eta_3) }{\eta\,\eta_1\eta_2\eta_3}\\
\times \delta_{1\eta^2}^{23} ff_{1}f_{2}f_{3}
\left(\tfrac{\eta^2}{f} + 
\tfrac{\eta_1^2}{f_1} -\tfrac{\eta_2^2}{f_2} -\tfrac{\eta_3^2}{f_3} 
\right)\mathrm{d}\eta_1\mathrm{d}\eta_2\mathrm{d}\eta_3\,,
\end{multline}
where  $f_i=f(\eta_i)$, $\eta_i=k_i/t^b$ for $i=1,2,3$, $f=f(\eta)$, and $\delta_{1\eta^2}^{23}=\delta(\eta^2+\eta_1^2-\eta_2^2-\eta_3^2)$.
The time dependence in the above equation must disappear, which implies $a=1/2$. Thus,
the equation for $f(\eta)$ becomes
\begin{multline} \label{eq:detadt}
 a f(\eta)+b\eta f'(\eta)\!\! =-2 \pi \int 
\frac{\min(\eta,\eta_1,\eta_2,\eta_3) }{\eta\,\eta_1\eta_2\eta_3} ff_{1}f_{2}f_{3} \\
\times \delta_{1\eta^2}^{23}  \left(\tfrac{\eta^2}{f}  +
\tfrac{\eta_1^2}{f_1} -\tfrac{\eta_2^2}{f_2} -\tfrac{\eta_3^2}{f_3}
\right)\!\mathrm{d}\eta_1\mathrm{d}\eta_2\mathrm{d}\eta_3\,. 
\end{multline}

Substituting the self-similar form \eqref{eq:self_dirc} into the definition of energy, we obtain
\begin{equation}\label{eq:Hnomega}
E= t^{3b-a} \int\limits_0^\infty \eta^2 f(\eta)\, \rmd\eta \,.
\end{equation}
Consider a temporal evolution of energy obeying the law $E(t) \propto t^{\lambda}$, with $\lambda =$~const $\ge 0$. Comparing it to \eqref{eq:Hnomega}, we obtain the scaling exponent $b=1/6+\lambda/3$.
Therefore, self-similar solution of the first kind is
\begin{equation}\label{eq:self-first}
    \nrad(k,t)t^{1/2}= f(k/t^b)\,\, {\rm with}\,\, b=1/6+\lambda/3\,.
\end{equation}
To characterize the temporal propagation of the direct cascade front, let us set a certain value $f_c$ such that $f_c \ll f_{max}$, where $f_{max}$ is the maximum value of function $f(\eta)$, find from the equation $f(\eta_{\rm cf})=f_c$ the value $\eta_{\rm cf}=\text{const}$, and define the location of spectral front of the direct cascade at a time moment $t$ as $\kcf(t)$ such that $\kcf(t)/t^b=\eta_{\rm cf}$.

Far behind the moving front, for $\eta\ll \eta_{\rm cf}$, we expect a power-law behaviour,
$f(\eta) \sim \eta^{-x}$ with $x\ge 0$. Substituting this power law into \eqref{eq:detadt}, we see that
in the limit $\eta \to 0$  each term of the left-hand side (LHS) is vanishingly small compared to the right-hand side (RHS) for $x>0$.
Therefore, for $x>0$
we conclude that for $\eta\ll \eta_{\rm cf}$ the spectrum tends to the solution of the equation RHS=0, i.e. to a  spectrum whose exponent $x$ is the same as one of the stationary solutions (but not the prefactor!). The borderline case $x=0$ is similar since it is the energy equipartition case for wich LHS=RHS=0. Thus, at $\eta\ll \eta_{\rm cf}$,  for the forced case we have  the direct cascade KZ exponent ($x=1$), whereas for the unforced case this should be the thermodynamic energy equipartition  
($x=0$). 
On the other hand, the unforced/undissipated case is tricky because the WKE blows up in a finite time $t^*$, whereas the self-similar solution is usually expected at large times. Nonetheless, we will see that the self-similar solution provides a reasonably good description of the long time evolution of the GPE spectrum (which, in contrast with the solution of WKE, does not blow up).

\subsection{The second-kind self-similarity for the inverse cascade} \label{2ndSS}

The inverse cascade has a finite capacity KZ spectrum, and therefore it is expected to exhibit a second-kind self-similarity in its dynamics.
This self-similar regime is characterized by the presence of a 
blowup time $t^*$ and it forms asymptotically very close to this time.
Note that the presence of forcing is unessential for this regime due to its self-accelerating blowup nature.
Previously, the self-similar solutions of the second kind of the
WKE associated with the GPE model were studied in \cite{semikoz1995kinetics,semikoz1997condensation,SemGreMedNaz,lacaze2001dynamical}, 
and its signatures were seen in the direct numerical simulations of the forced/dissipated 3D GPE in 
\cite{shukla2022nonequilibrium}, 
and in unforced simulations from
\cite{zhu2022testing}, with  prior attempts made in \cite{berloff2002scenario, during2009breakdown}. In the present paper, we will recover the previous results and complete them with more detailed considerations of both forced and unforced systems, as well as by considering the setups in which the inverse and the direct cascades show up simultaneously.

For the second-type self-similarity, we assume the following form of the spectrum:
\begin{equation}\label{eq:self_inv}
\nradkt=g(\xi)\tau^{-r}\,\,\text{with}\,\xi=k/\tau^{m}\,\,\text{and}\, \tau=t^*-t,
\end{equation}
where $t^*$ is the blowup time.
Substituting \eqref{eq:self_inv} into \eqref{eq:WKE1}
and requiring that the resulting equation involves only the 
similarity variable $\eta$ and not $\tau$, we get  $r=1/2$ and
\begin{multline}\label{eq:dxidt}
r g(\xi)+m\xi g'(\xi) =
2\pi\medint\int  \frac{\min(\xi,\xi_1,\xi_2,\xi_3)}{\xi\,\xi_1\xi_2\xi_3}
gg_{1}g_{2}g_{3}  \\
\times \delta_{1\xi^2}^{23}
\left(\tfrac{\xi^2}{g} + 
\tfrac{\xi_1^2}{g_1} -\tfrac{\xi_2^2}{g_2} -\tfrac{\xi_3^2}{g_3} 
\right)\mathrm{d}\xi_1\mathrm{d}\xi_2\mathrm{d}\xi_3\,,
\end{multline}
where  $g_i=g(\xi_i)$, $\xi_i=k_i/\tau^m$ for $i=1,2,3$, $g=g(\xi)$,
and $\delta_{1\xi^2}^{23}=\delta(\xi^2+\xi_1^2-\xi_2^2-\xi_3^2)$. 
It was shown numerically  in \cite{semikoz1995kinetics,lacaze2001dynamical} and proven analytically in
\cite{SemGreMedNaz}  that $g(\xi) \propto \xi^2$ for $\xi\ll1$ which corresponds to the thermodynamic energy equipartition spectrum. For $\xi \gg 1$, the spectrum approaches a power law 
$g(\xi) \propto \xi^{-x^*}$   with exponent $x^*=1/(2m)$,
which is anomalous, i.e. neither KZ nor thermodynamic. 
This exponent has been numerically explored in several papers by simulating the WKE evolution, seeking for $n_{\omega} \sim \omega^{-x}$, where $x=1+x^*/2$. The references \cite{semikoz1995kinetics, semikoz1997condensation} reported a value of $x$ as $1.24$ ($x^*= 0.48$), while in the reference \cite{lacaze2001dynamical}, a value of  $x=1.234$ ($x^* \approx 0.47$) was obtained. More recently, 
solving directly the nonlinear eigenvalue problem associated with \eqref{eq:dxidt} in \cite{SemGreMedNaz}, the most carefully determined candidate values were found to be $x=1.22$  and $x=1.24$, corresponding to $x^*=0.44$ and  $x^*=0.48$, respectively.

Note that  our self-similar solution of the second kind
\begin{equation}\label{eq:self-second}
    \nrad(k,t)\tau^{1/2}= g(k/\tau^m)\,\, {\rm with}\,\, m=1/(2x^*)\,
\end{equation}
implies that the spectra $\nradkt$ for various time moments collapse into a single curve $g(\xi)$ when the time is close to $t^*$. 
Substituting the established behavior $g(\xi) \propto \xi^2$ as $\xi\to 0$, which was proved in \cite{SemGreMedNaz} (based on the nonlocality of interaction of scales $\xi \ll 1$ with scales $\xi \sim 1$), into \eqref{eq:self-second}, we can deduce that
the quantity
\begin{equation}\label{eq:self-G}
G(\tau)= \lim_{k\to 0} \nrad(k,t) \tau^{1/2+2m}/k^2\end{equation}
must tend to a constant as $t$ is approaching $t^*$ from below ($\tau \to +0$). We shall use this as one of the tests of self-similarity in our numerics.

\subsection{The third-kind self-similarity for the inverse cascade}

As we explained above, the first- and the second-kind self-similarities are different because in the former case the stationary spectrum is formed right behind the propagating front, whereas for the latter case an anomalous power-law spectrum forms for $t\to t^*$ ($t<t^*$). The anomalous power law is further replaced by a stationary spectrum---the process that takes the form of a reflected wave propagating back from the dissipation wave number to the forcing one. This new type of self-similar behaviour was first studied for the direct cascade systems in \cite{Bell_2018,Nazarenko2019}, but it is natural to expect it for all finite-capacity systems, in particular, for the BEC WT inverse cascade considered in the present paper. This behaviour does not fit the Zeldovich's  first/second-kind classification and, therefore, was named 
the third-kind self-similarity in \cite{Bell_2018,Nazarenko2019}.

The third-kind self-similarity is realized for $t\to t^*$ ($t>t^*$); it is characterised by the spectrum
\eqref{eq:self_inv} in which now $\tau =t- t^*>0$. In the present paper, we will not study such a behaviour in detail because the numerical resolution of our simulations is insufficient for making definitive conclusions. However, we will comment on the signatures that are consistent with the reflected wave scenario in the results of the WKE numerics and on the absence of such signatures in the GPE numerics.

\subsection{Free decay: blowup vs no-blowup initial data}
\label{E/N}

As mentioned in section \ref{2ndSS}, the second-kind self-similarity is observed for the WKE in the inverse-cascade settings, both with and without forcing. This behaviour is a precursor to the condensation at $k=0$ which sets in at a finite time $t^*$. Actually, for unforced systems, this kind of evolution is expected only for sufficiently low-$k$ initial data, as follows from the standard Einstein's condensation argument applied to the classical waves \cite{Connaughton2005}. 
This argument consists in a statement that Bose-Einstein condensation occurs when no equilibrium Rayleigh-Jeans (RJ) spectrum \eqref{eq:thermal} $\nradk =4\pi k^2T/(k^2 +\mu)$ (with 
$T,\mu =$const) can be found containing the same amount of $N$ and $E$ as in the initial condition. (Note that it is necessary to assume that the system is truncated at the UV-cutoff $k_{max}$ in order to make $E$ and $N$ finite.) Specifically, the minimal possible value of $E/N$ in the RJ spectrum occurs for $\mu=0$, and this value is equal to $k^2_{max}/3$.
Of course, this argument implies that the thermal equilibrium is an attracting state, i.e. that there is a mixing mechanism leading to relaxation to this state either due to a coupling to a thermal bath (not our case) or provided by nonlinear wave interactions (our case). In the latter case, the required mixing may be absent for certain special initial data and the WKE-governed system goes, e.g., through a periodic evolution~\cite{escobedo2015theory}. 

The previous discussion implicitly assumes that energy and total number of particles are conserved during the temporal evolution, even in the presence of an UV-cutoff.  At first sight, it seems contradictory that a truncated system exhibiting a direct cascade could conserve energy, as one might expect naively that interacting wave modes will excite wave numbers beyond $k_{\max}$, creating an energy leakage. However, by truncating a system, one actually kills such interactions so that the cascade can not pass through $k_{\max}$. 

The introduction of an UV-cutoff in a non-linear partial differential equation allows for a simple statistical mechanics description of thermal states, as first realised by T.D. Lee and R. Kraichnan for the truncated Euler equation \cite{Lee_StatisticalPropertiesHydrodynamical_1952,Kraichnan-inertial}. Kraichnan introduced the concept of absolute equilibria in which Fourier modes of the velocity field are in thermal equilibrium and obey Gibbs statistics, i.e they have the distribution $\propto\exp{[- E/T]}$, where $E$ is the kinetic energy of the flow. The relaxation towards thermal equilibrium presents a rich and complex dynamics which exhibits turbulent cascades prior to complete thermalisation \cite{Cichowlas_EffectiveDissipationTurbulence_2005}. This process was later extended to the cases of helical flows \cite{Krstulovic_CascadesThermalizationEddy_2009}, the Burgers equation \cite{Ray_ResonancePhenomenonGalerkintruncated_2011}, magnetohydrodynamics \cite{Krstulovic_AlfvenWavesIdeal_2011}, and in particular, to the case of the truncated Gross-Pitaevskii equation \cite{KrstulovicDispersiveBottleneck,KrstulovicTruncatedGP}. Note that the RJ spectrum \eqref{eq:thermal} is an absolute equilibrium of the truncated GP system for small amplitude waves. 
In the context of finite temperature BECs, the truncation of the GP equation is justified by a semi-classical approximation in which only particles with momenta such that $\hbar\omega(k_{max})\le k_{\rm B}T$ are taken into account (with $T$ the temperature and $k_{\rm B}$ and $\hbar$ the Boltzmann constant and Planck constant). In the context of weakly nonlinear optical waves propagating in a multimode fiber, the frequency cutoff naturally arises from experimental conditions \cite{Baudin2020Condensation}.
We introduce in the following the truncated GP and properly define the truncated wave kinetic equation.

\subsubsection{Truncated Gross-Pitaevskii equation.}

The truncated GP equation is easily defined from the standard GP equation expressed in Fourier space. It results from a Galerkin projection at the wave number $k_{max}$ and reads
\begin{equation}
  \frac{\partial\hat{\psi}_{\bf k}  }{\partial t}=-ik^2\hat{\psi}_{\bf k}  - \sum_{\bf 123}\theta_k\theta_1\theta_2\theta_3\hat{\psi}_{\bf 1}^* \hat{\psi}_{\bf 2}\hat{\psi}_{\bf 3}\delta^{01}_{23},
 \label{eq:TGPE}
\end{equation} 
where $\theta_k=1$, if $|{\bf k}|\le k_{max}$ and $\theta_k=0$ otherwise. It can be easily seen that this equation derives from the Hamiltonian $H=\sum_{|\bf k|\le k_{\rm max}}k^2|\hat{\psi}_{\bf k}|^2+\sum_{\bf 1234}\theta_1\theta_2\theta_3\theta_4\hat{\psi}_{\bf 1}^*\hat{\psi}_{\bf 2}^* \hat{\psi}_{\bf 3}\hat{\psi}_{\bf 4}\delta^{12}_{34}$. 
The truncated  Hamiltonian preserves invariance with respect to the time and phase shifts  and, therefore, Eq.~\eqref{eq:TGPE} also conserves the total energy and the total number of particles.

The truncation of the Gross-Pitaevskii model was first explicitly introduced by Davis \etal \cite{Davis_SimulationsBoseFields_2001} to study the process of condensation in Bose gases by performing direct numerical simulations of the GP equation. Since then, the truncated (or sometimes called projected) GP equation has become an important model for finite temperature BECs. It has been used for studying the interaction of vortices and particles with thermal waves \cite{Krstulovic_AnomalousVortexringVelocities_2011,Giuriato_StochasticMotionFinitesize_2021}, as well as the quantum turbulence at finite temperatures \cite{Shukla_GPViscosity}. Note that RJ spectrum \eqref{eq:thermal}, and the argument given for classical condensation of waves in \cite{Connaughton2005} is only qualitative in the case of infinite systems. Indeed, close to the condensation transition, the system becomes fully non-linear at very low wavenumbers and the WT theory can not be applied there. The whole Hamiltonian \eqref{eq:cons-H} should be taken into account, which corresponds to the energy of the well-known $\lambda-\phi^4$ theory, describing second-order phase transitions \cite{KrstulovicDispersiveBottleneck}.
On the other hand, in finite (e.g., trapped) systems, the lowest wavenumber is finite, and in principle, the system may remain weakly nonlinear even when all the particles condense at the lowest energy level.

\subsubsection{Truncated wave kinetic equations\label{subsubsec:truncatedWKE}}

\begin{figure}[h!]
    \begin{center}
    \centering{\includegraphics[scale=0.8]{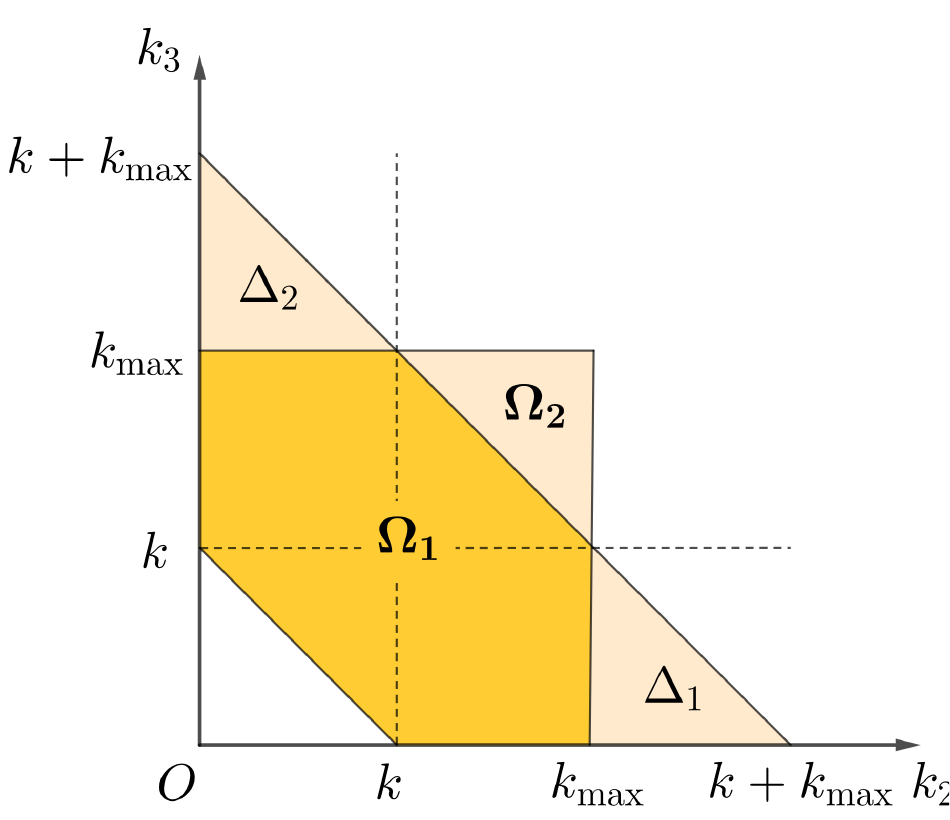}}
    \caption{Domain of integration of the collision term of truncated WKE.
    }\label{IntDomain}
    \end{center}
    \end{figure}

The truncated WKE immediately follows from Eq.~\eqref{eq:TGPE} and its Hamiltonian, as the truncation can be interpreted as a collisional matrix. It simply reads
\begin{multline}\label{eq:TWKE}
  \frac{\partial \nrad_k}{\partial t}= 
  2\pi \medint\int \theta_k\theta_1\theta_2\theta_3
  \frac{\min\left(k,k_1,k_2,k_3\right)}{k\,k_1k_2k_3}
n_{k}^{\rm rad} n_{{k_1}}^{\rm rad}n_{{k_2}}^{\rm rad}n_{{k_3}}^{\rm rad} \qquad \\
 \times \delta(\omega^{01}_{23}) \left(\tfrac{k^2}{n_{k}^{\rm rad}} + \tfrac{k_1^2}{n_{{k_1}}^{\rm rad}} - \tfrac{k_2^2}{n_{{k_2}}^{\rm rad}}-\tfrac{k_3^2}{n_{{k_3}}^{\rm rad}}\right)  \mathrm{d}k_1\mathrm{d}k_2\mathrm{d}k_3\,,  
\end{multline} 
  which also naturally conserves the truncated invariants. 
The truncation acts on all wavenumber $k$, $k_2$, $k_3$ and $k_1=k-k_2-k_3$, which reduces the integration domain to the area $\Omega_1$ in Figure~\ref{IntDomain}. 
Note that performing a naive truncation on the WKE, which could be for instance keeping the domains $\Delta_1$, $\Delta_2$ and $\Omega_2$, leads to an energy leak through $k_{\rm max}$. The spurious consequence of such a choice is discussed in \ref{App:Naivetruncation}.

More interesting, it is well known that the WKE admits an $H$-theorem which states that the entropy $S=\int \log{n_{\bf k}}\mathrm{d}{\bf k}$ grows monotonically in time \cite{ZLF}.
In particular, the theorem holds in the truncated case with the entropy truncated at $k_{\rm max}$. Moreover, note that the RJ spectra maximize the entropy at fixed total energy and number of particles. Therefore, it is natural to assume that in the absence of finite-time blowup, the truncated WKE solutions will tend to the RJ solution if temperature and chemical potential can be determined.

Having introduced the truncated WKE, we can formulate the following proposition for the freely decaying WKE systems with a UV-cutoff $k_{max}$.
(i) For initial data such that $E/N \geq k^2_{max}/3$, the finite time blowup at $k=0$ is absent, and, respectively, no self-similar evolution of the second kind is observed. (ii) Assuming additionally that the initial data is not pathological and there is a suitable mixing present in the system (validity of the $H$-theorem), the spectrum asymptotes to the RJ spectrum for $t\to \infty$. Conditions for suitable mixing are to be determined. (iii) For initial data with $E/N < k^2_{max}/3$, the finite time blowup will occur at $k=0$ (provided a suitable mixing condition) and, respectively,  self-similar evolution of the second kind is observed in the vicinity of the blowup time. 

Note that the finite-time blowup  for a specific initial condition was rigorously proven in Ref~\cite{escobedo2015theory}. In their case, there was no maximum wave number (formally, $k_{max}=\infty$), so the blowup condition specified above in (iii) was satisfied.

For the forced systems, there will be a continuous supply of particles, and we can conjecture that the WKE system will always blow up in a finite time (in the absence of dissipation at the low-$k$ region) irrespective of the initial condition and the forcing strength.

\section{Numerical results}

\subsection{Numerical setup}

To study the self-similar behavior of the different kinds, we consider two types of setups: the unforced (free-decay) and the forced systems. We further subdivide the forced setups into the direct and the inverse cascades---with forcing at low and at high wave numbers, respectively. Moreover, we consider the inverse cascade in two cases exhibiting qualitatively different behaviors ---with and without dissipation at the smallest wave numbers. The latter case leads to condensation.  

We numerically simulate GPE \eqref{eq:GPE} and WKE \eqref{eq:WKE}.
The GPE is solved by using the standard massively-parallel pseudo-spectral code FROST \cite{KrstulovicHDR} with a fourth-order Exponential Runge-Kutta temporal scheme (see \cite{zhu2022testing}). We discretize the  $L^3$-periodic box using $N_p^3$ collocation points. 
The WKE is solved 
using the code developed in \cite{SemGreMedNaz,zhu2022testing,zhu2022direct}. This code uses a decomposition of the integration domain in the RHS of \eqref{eq:WKE}, so that in each subdomain the integrand is a highly-smooth function.
We simulate the WKE in the interval $\omega\in[\omega_{\rm min},\omega_{\rm max}]$, and we set $n_{\omega}= n_{\omega_{\rm min}}$ for $\omega<\omega_{\rm min}$, and $n_{\omega}=0$ for $\omega>\omega_{\rm max}$. 
The second-order Runge-Kutta scheme is employed to march the time for free-decay cases, and a new approach inspired by Chebyshev interpolation and schemes described in \cite{Sem} is used for the time integration in the forced cases.
The WKE solvers employed in \cite{zhu2022testing, zhu2022direct}
attained superior resolution and broader computing ranges while demanding significantly fewer computing resources  than the GPE solver.

For GPE simulations, the spherically-integrated wave-action spectrum is computed as
\begin{equation}
n^{\text{rad}}(k,t)=\frac{1}{D_k} \sum_{{\bf k}\in \Gamma_k } |\hat{\psi}({\bf k},t)|^2 \,,
\label{eq:GPE1D}    
\end{equation} 
where  $\Gamma_k$ is the spherical shell around $|{\bf k}|=k$ with thickness  $D_k$.
In WKE simulations, $n^{\text{rad}}(k,t)=4\pi \omega n_{\omega}(t)$.

In all direct-cascade simulations, the propagation of the spectral front $\kcf(t)$ is measured by setting a small threshold value $\varepsilon$ and finding for different $t$ the value $\kcf$ such that $\nrad(\kcf,t)=\varepsilon.$

\subsection{Free-decay simulations for low $E/N$}

For the free-decay setup, we will, first of all, consider  cases which
(according to our conjecture in section \ref{E/N}) satisfy the finite-time blowup condition for WKE, $E/N < k^2_{max}/3$. We start 
simulations of these cases with  Gaussian-shaped   spectra, similar to the ones used in \cite{zhu2022testing}:
\begin{equation}
\label{initialData}
n^{\text{rad}}(k,0) = g_0\exp\biggl(\frac{-(k-k_s)^2}{\sigma^2}\biggr).
\end{equation}
We present three free-decay simulations (case~1, case~3, and case~4) with the parameters shown in Table \ref{tab:free}.  
\begin{table}[h!]
\begin{center}
\begin{tabularx}{0.47\textwidth}{|X|X|X|X|X|X|X|}
\hhline{|=|=|=|=|=|=|=|}
case & model & $\omega_{\rm min}$ & $\omega_{\rm max}$ & $k_s$ & $g_0$  & $\sigma$  \\
\hline
1  & WKE  & $10^{-5}$ & 10 & 1.5 & 1 & 0.2 \\
2  & WKE  & $10^{-2}$ & 86 & 55 & 1 & 2.5 \\
\hhline{|=|=|=|=|=|=|=|}
case & model & $L$ & $N_p$ & $k_s$ & $g_0$  &  $\sigma$ \\
\hline
3& GPE  & $8\pi$  & 720 & 1.5  & 1 & 0.2 \\
4& GPE  & $8\pi$ &  512  & 22 & 1 & 2.5  \\
5& GPE  & $8\pi$ &  512  & 35 & 2 & 1  \\
\hline
\end{tabularx}
\caption{\label{tab:free}
Numerical parameters for the free-decay simulations.
}%
\end{center}
\end{table}

\begin{figure*}[hbtp]
\begin{center}
\includegraphics[scale=1]{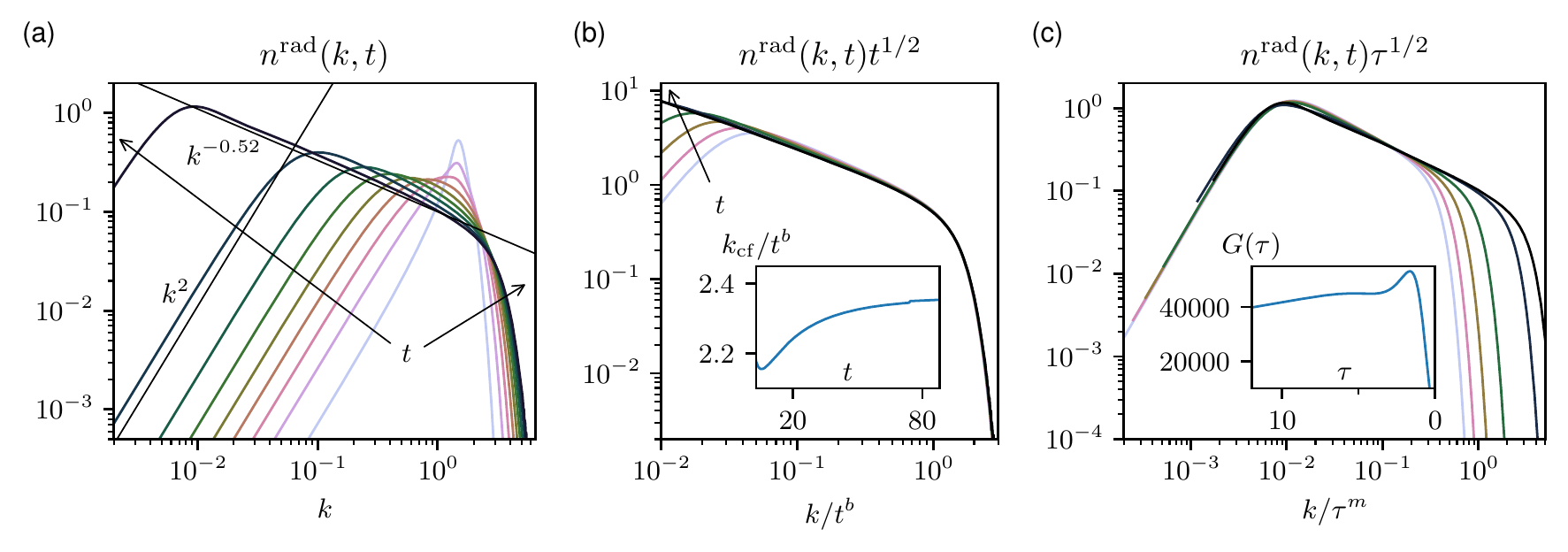}
\caption{Numerical results obtained in WKE simulation (case~1) for free-decay case with $k_s=1.5$: (a) Spherically-integrated wave-action spectrum $\nrad (k,t)$ for the times $t=10\,,20\,,30\,,40\,,50\,,60\,,70\,,80\,,88.5$; (b) $\nrad (k,t)$ compensated by $t^{1/2}$ vs. compensated wave number $k/t^b$, and inset for the time evolution of the compensated wave front $\kcf/t^b$; (c) $\nrad (k,t)$ compensated by $\tau^{1/2}$ vs. compensated wave number $k/\tau^m$, and inset for $G(\tau)$ for the time close to blowing up. Spectra in panels (b) and (c) are given for  $t=80\,, 82\,, 84\,, 86\,,88\,, 88.5$. }\label{fig:nk_free_dirc_wke}
\end{center}
\end{figure*}
Figure \ref{fig:nk_free_dirc_wke} is obtained by performing WKE simulation using the parameters of case~1 described in Table.~\ref{tab:free}. The initial spectrum was centered at $k_s=1.5$. We took essential care of setting the distribution of the $k$-grid points in order to guarantee high accuracy at both small and large $k$'s.  
Figure \ref{fig:nk_free_dirc_wke}(a) shows the dual cascade of wave-action $\nrad(k,t)$ for the values of time $t\in [10,88.5]$.
The front of $\nrad(k,t)$ propagating to the left develops a thermal particle-equipartition scaling  $\sim k^2$, as was also discussed in \cite{SemGreMedNaz, zhu2022testing}. At $t=88.5$ -- close to the blowup time $t^*\approx 89.5$ -- the scaling $\sim k^{-0.52}$  is seen between the front propagating to the left and the initial peak. 
 For late pre-blowup times  $t\in [80,88.5]$, in Figure \ref{fig:nk_free_dirc_wke}(b) we plot self-similar solutions of the first kind given by \eqref{eq:self-first}  with $\lambda=0\,,b=1/6$ (corresponding to the free-decay case), i.e. we plot $\nrad(k,t)t^{1/2}$ vs. $k/t^{1/6}$. Respectively, for the same period of time, in Figure \ref{fig:nk_free_dirc_wke}(c) we show a self-similar solution of the second kind given by \eqref{eq:self-second}  with $x^*=0.52$ and $t^*=89.5$ (i.e. $\nrad(k,t)\tau^{1/2}$ vs. $k/\tau^{1/1.04}$).
Inset in  Figure \ref{fig:nk_free_dirc_wke}(b) presents the time evolution of the compensated  spectral wave-action front $\kcf/t^b$ in the direct cascade setting. 
The fact that $\kcf/t^b$ tends to be a constant at late times is in almost perfect agreement with the prediction of first-kind self-similarity with $b=1/6$.
On the other hand, Figure \ref{fig:nk_free_dirc_wke}(c) shows a perfect collapse  of plots $\nrad(k,t)\tau^{1/2}$ vs. $k/\tau^m$ onto the same curve in the inverse cascade setting. Moreover, in the interval $2 < \tau < 7$ we obtain almost constant $G(\tau)$ as predicted by \eqref{eq:self-G} for small $\tau$, see the inset of Figure \ref{fig:nk_free_dirc_wke}(c). This dependence breaks down for very small $\tau$ because of the presence of the minimal frequency $\omega_{{\rm min}}$ and a small uncertainty in finding $t^*$ in numerics. 

It is remarkable that we can see self-similarities of both the first and the second kinds in one single WKE simulation. This is mostly because of the efficient numerical method that allows adaptive $k$- (or $\omega$-) space discretization with great accuracy. However, the blowup phenomenon in WKE evolution means that one cannot continue the WKE solution beyond $t=t^*$, whereas formally the self-similar solution of the first-kind \eqref{eq:self-first} is expected asymptotically as
$t\to \infty$.
Taking into account that the direct cascade propagation in the free-decay situation is quite slow ($\kcf \sim t^{1/6}$), it is worth noting that one can observe only the tendency of $\kcf/t^b$ to a constant, see the inset of Figure \ref{fig:nk_free_dirc_wke}(b). Even though, it is surprising to see
  that the self-similar behavior in Figure \ref{fig:nk_free_dirc_wke}(b) develops relatively early in time.

\begin{figure}[htbp]
\begin{center}
\includegraphics[scale=1]{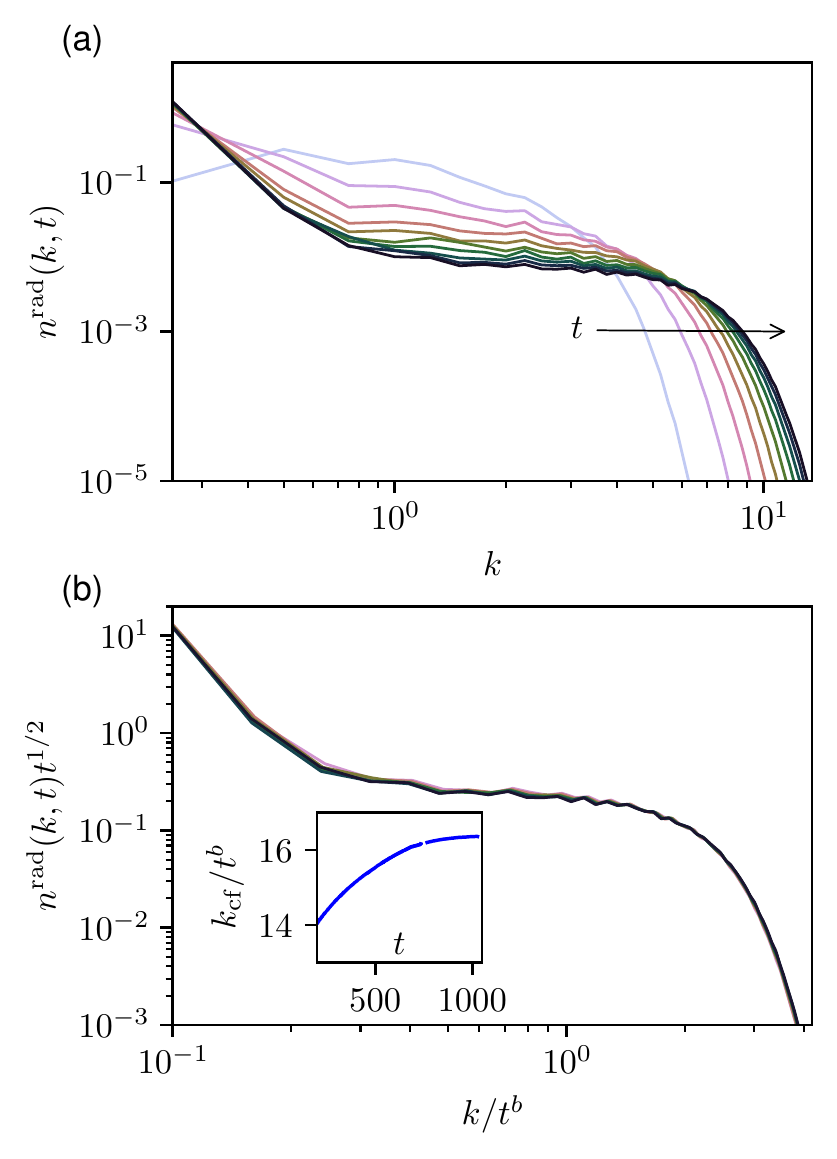}
\caption{Numerical results for free-decay GPE simulation (case~3) with $k_s=1.5$. (a) Spherically-integrated wave-action spectrum $\nrad (k,t)$ for times $t=100\,,200\,,300\,,400\,,500\,,600\,,700\,,800\,,900\,,1000$; (b) $\nrad (k,t)$ compensated by $t^{1/2}$ vs. compensated wave number $k/t^b$ for times $t=850\,,880\,,910\,,940\,,970\,,1000$ and inset for the time evolution of the compensated wavefront $\kcf/t^b$. }\label{fig:nk_free_dirc_gpe}
\end{center}
\end{figure}

Simulating GPE is more challenging than WKE, and presently it is impossible to include a sufficiently wide range of scales for observing self-similarities of both the first and the second kind simultaneously in one numerical run. Thus, we ran two different setups with the initial spectrum in the low and high $k$'s in order to observe the first- and the second-kind self-similarities in the direct and the inverse cascades, separately.
To implement a direct cascade for the  free-decay case, we simulate GPE (case~3 in Table \ref{tab:free}) using the same initial $\nrad(k,0)$ as in case~1. However, the GPE discretization is much coarser than the one used for WKE, so the initial data is in the $k$'s where the discreteness is substantial (with the maximum at the wave number equal to the six wave number spacings $2\pi/L$).
For this precise reason the inverse cascade evolution is suppressed from the beginning, and the direct cascade dynamics can be viewed as post-$t^*$. Note that we add friction $D_{\bm 0}=10^3$ at $k=0$ (see section~\ref{sec:forcing} for the definition of $D_{\bm 0}$) to effectively prevent the condensation.
It should be emphasized that $D_0$ needs to be sufficiently large, typically $D_0 \ge 10^3$ to guarantee the condensate rate $\nrad(0,t)/N < 10^{-4}$ across all relevant simulations.
After a small initial change ($\sim 0.1\%$ for $0<t<150$) the  relative deviations of the particle and energy densities $N$ and $H$ remain constant with accuracy $\sim 0.01\%$, which means that the choice  $\lambda=0$ in \eqref{eq:self-first} is still suitable. 
We present the time evolution of wave-action spectra and the  self-similar functions of first-kind  obtained in this simulation 
 in panels (a) and (b) of Figure~\ref{fig:nk_free_dirc_gpe}, respectively, and in the inset of the panel (b) -- time evolution of the compensated spectral front for the direct cascade.  At late times, thermal-like equilibrium energy equipartition  scaling $\sim k^0$ is observed at the intermediate $k$-range (between the low-$k$ condensate and the high-$k$ spectral front) as predicted. Also as predicted by the first-kind self-similarity, we observe  the collapse  of  curves $\nrad(k,t)t^{1/2}$ vs. $k/t^b$ for different $t$'s, and we see an asymptotic tendency of $\kcf/t^b$ to a constant. 
It is noteworthy that the recent experiment \cite{glidden2021bidirectional} 
 achieved  dual cascades with almost constant $N$ and $H$,  reporting a value of $b \approx 0.14$. This finding closely aligns to our prediction $b=1/6$.
Note that even though the GPE  and the WKE runs (cases~3 and~1, respectively) share the same initial spectrum, the results obtained by these two simulations deviate quickly because of totally different discretization in $k$-space. The GPE simulation bypasses the pre-$t^*$ evolution, whereas the WKE evolution ends at $t^*$. In principle, one could regularise the WKE for describing the post-$t^*$ evolution by coupling it to an equation for the condensate mode $k=0$, as it is done in \cite{semikoz1995kinetics,semikoz1997condensation},
but studying the resulting equation is beyond the scope of the present paper. It is interesting, however, that the energy equipartition spectrum observed in our GPE simulation was also claimed to be relevant to the post-$t^*$ evolution of the regularised WKE-condensate system in \cite{semikoz1995kinetics,semikoz1997condensation}.

\begin{figure}[htbp]
\begin{center}
\includegraphics[scale=1]{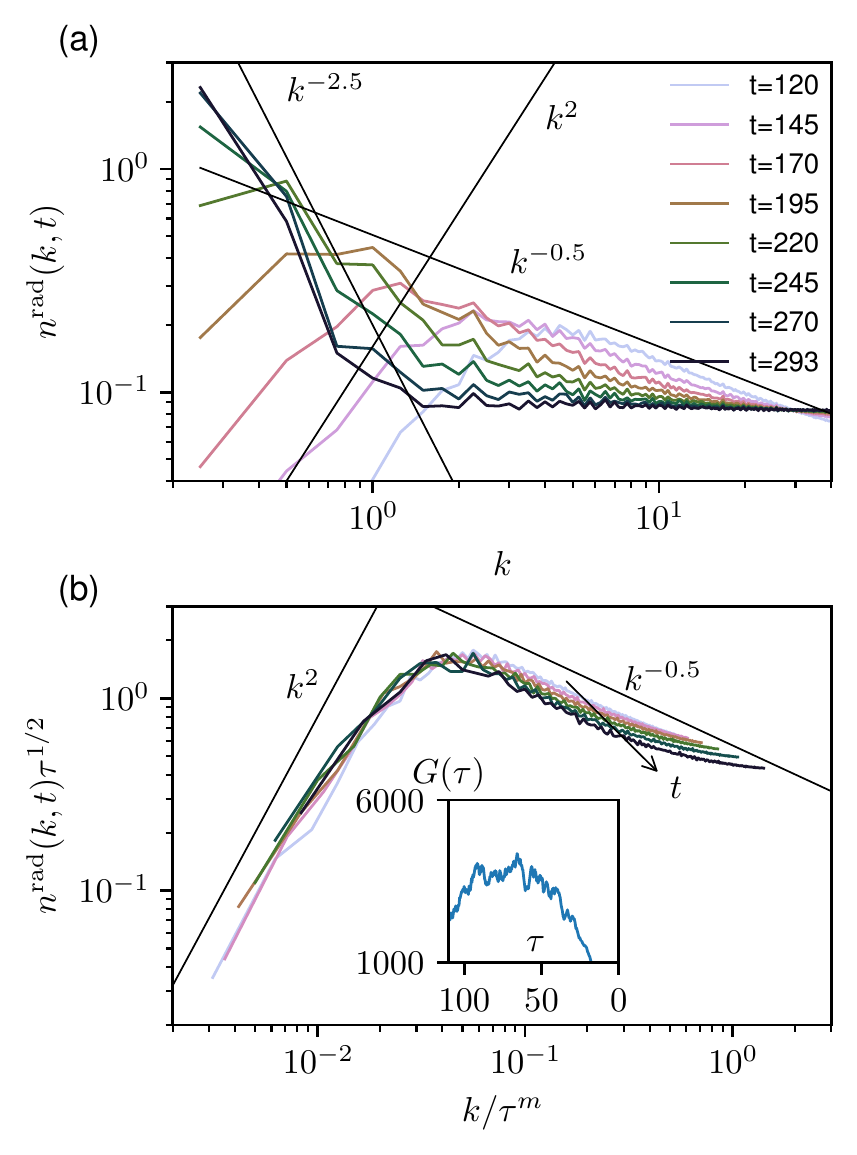}
\caption{Numerical results for free-decay GPE simulation (case~4) with $k_s=22$: (a) Evolution of the spherically-integrated wave-action spectrum $\nrad (k,t)$;  (b) the spectrum $\nrad (k,t)$ compensated by $\tau^{1/2}$ vs. compensated wave number $k/\tau^m$ for $t=120\,,130\,,140\,,150\,,160\,,170$. Inset: $G(\tau)$ for the time close to the blowup.}\label{fig:nk_free_inv_gpe}
\end{center}
\end{figure}
The second kind of self-similarity is also studied via GPE free-decay simulations with an initial spectrum centered at relatively large wave number $k_s=22$ (which is yet low enough for the WKE blowup condition, $E/N < k^2_{max}/3$); see case~4 in Table~\ref{tab:free}. 
Figure \ref{fig:nk_free_inv_gpe}(a) presents the late time evolution of $\nrad(k,t)$ for  $t\in [120\,,293]$ (early time behavior was discussed in \cite{zhu2022testing}). One can see clearly the second kind  self-similarity for $t\in[120\,, 170]$ with $k^2$-scaling at low $k$'s and $k^{-0.5}$-scaling at large $k$'s. The shape of $\nrad(k,t)$ starts to change after $t=170$  showing rapid accumulation of waves in the smallest $k$ (condensation). 
We observe
\
and quasi-thermal (energy-equipartition) constant spectra at high $k$'s.
Similar behavior was reported in \cite{shukla2022nonequilibrium} for the GPE simulations  with forcing
at large $k$'s. This confirms that the forcing is not important for the self-similarity of the second kind because the characteristic time associated with the forcing is much greater than the characteristic nonlinear time near  $t^*$.
We estimate the dependencies of $\nrad(k,t)\tau^{1/2}$ on $k/\tau^m$ with $x^*=0.5$ by using  \eqref{eq:self-second} for each value of $t^*$ in a finely girded range of $(170, 230]$.
This investigation focuses on the data within the established self-similar regime.
The value $t^*\approx 200$ minimizes the least-square deviations of the data during the intermediate self-similar phase ($t\in[120,170]$).
Figure \ref{fig:nk_free_inv_gpe}(b) plots the self-similar functions $\nrad(k,t)\tau^{1/2}$ vs. $k/\tau^m$ for $t\in[120\,,170]$. A visible collapse of curves is observed except for high $k$. We also find a relatively constant $G(\tau)$ for $30\le \tau \le 100$ in the inset, which coincides with the self-similarity for $t\in[120,170]$. Note that, unlike WKE, the GPE evolution does not lead to a blowup, and the self-similarity of the second kind is observed as an approximate intermediate asymptotic only. This is natural because the WKE fails to be applicable while approaching $t^*$ both because of the rapid nonlinearity growth and the increased sharpness of the spectrum at low $k$'s where the discreteness of the $k$-space is essential. 
This results in the long-term deceleration of  GPE dynamics compared to WKE.
For the GPE we should also note that the condensate does not occur at the mode $k=0$ only, like in WKE, but takes a form of a sharp spectrum at low $k$'s.
We suggest a scaling of $\sim k^{-2.5}$ as a visual guide, obtained by fitting the data on a limited number of grid points, but this scaling is probably not universal.

\subsection{Free-decay simulations for high $E/N$}

Now, let us consider  cases which
(according to our conjecture in section \ref{E/N}) satisfy the  no-blowup condition for WKE, $E/N \geq k^2_{max}/3$.
We take initial conditions as case~2 in Table~\ref{tab:free} for WKE simulation, and as case~5 for GPE simulation, respectively.

\begin{figure*}[htbp]
\begin{center}
\centering{\includegraphics[scale=1]{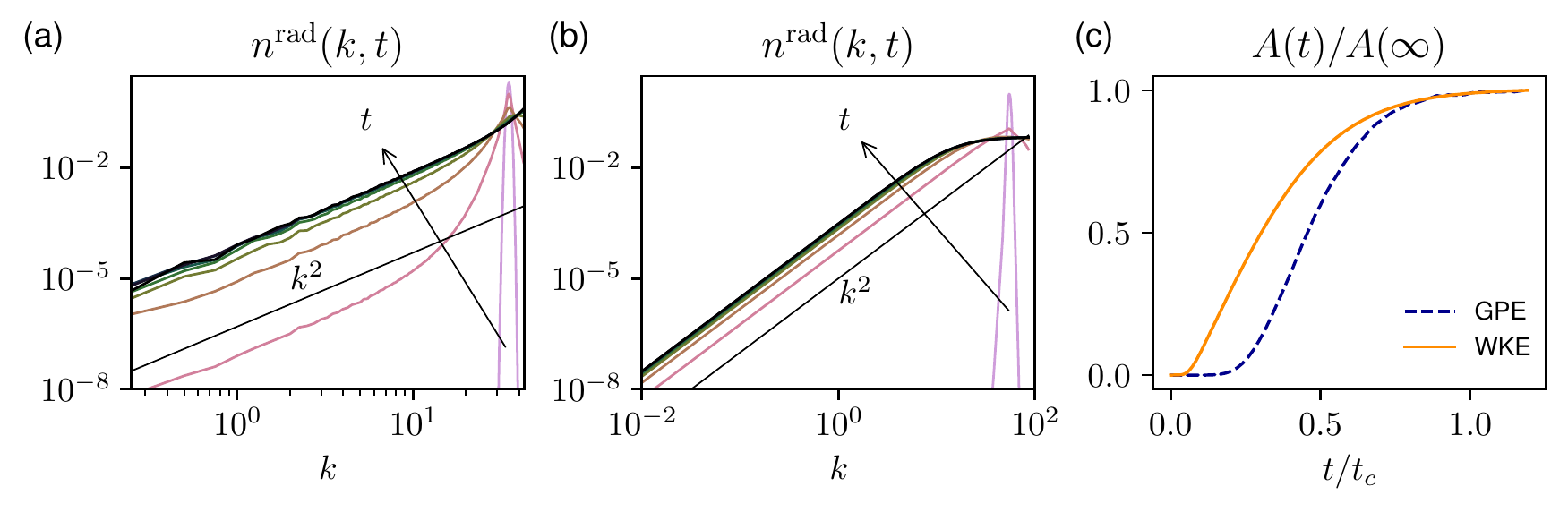}}
\caption{Numerical results with large $E/N$: (a)  Spectrum $\nrad (k,t)$ in GPE simulation (case~5) for  times $t=0, 40, 80, 120,...,320$;  (b) Spectrum $\nrad (k,t)$ in the WKE simulation (case~2) for  times $t=0, 250, 500, 750,...,2000$; (c) Values of $A(t)/A(\infty)$ for the prefactor $A(t)$ in the fit $\nrad (k,t) = A(t) \, k^2$ for both cases.}\label{fig:largeE/N}
\end{center}
\end{figure*}
The simulation results are presented in Figure \ref{fig:largeE/N}. The time evolution of spectra $\nrad (k,t)$ obtained 
 by solving GPE are given in panel (a) for $t\in[0,320]$, and the results obtained by solving WKE are given in panel (b) for $t\in[0,2000]$. 
For GPE simulation, as predicted in section \ref{E/N}, we see neither a tendency to low-$k$ condensation
nor any signatures inherent to self-similar behavior of the second kind. The spectrum quickly takes form $\nrad (k,t) = A(t) \, k^2$ in a widening $k$ range, with
$A(t) \to \,$const for $t\to \infty$, as shown in Figure \ref{fig:largeE/N} (c) ($t_c$ is defined as the time such that $A(t_c)$ has 1\% deviation from $A(\infty)$). In other words, the spectrum asymptotes to the (stationary) RJ spectrum with $\mu/T \to \infty$ which is a  thermodynamic state describing the particle equipartition. Note that the particle equipartition is realized instead of any other spectrum in the RJ family because it corresponds to the chosen initial condition with $E/N$ close to its maximum possible value $k^2_{max}$.
Similarly, the truncated WKE evolution in Figure \ref{fig:largeE/N} (b) and (c) also exhibits the same behavior as the GPE with a clear thermalization to the RJ spectrum. 

The case with a naive truncation in which energy leaks through the cutoff $k_{max}$ is discussed in Appendix~\ref{App:Naivetruncation}. In short, this leakage leads to a decreasing value of $E/N$, such that even if initially it is larger than $ k_{max}^2/3$, after some time this condition is violated followed by a condensation-type blowup in finite time.

\subsection{Simulations with forcing}\label{sec:forcing}

Unlike the free-decay case, all the simulations with forcing start with a zero initial field. In GPE simulations, we add the forcing term $F_{\bf k} (t) $ and the dissipation term $-D_{\bf k}\widehat{\psi}_{\bf{k}} (t) $ to the Fourier transform of the RHS of GPE \eqref{eq:GPE}. 
In the GPE simulation for the direct cascade (case~6 in Table \ref{tab:forcing},
we add forcing by fixing $\nrad(k,t)=n_{\rm f}$ on a  spherical  shell in $k$-space with $0.5 \le k \le k_{\rm f}$, whereas for the inverse cascade (cases~6 and~7 in Table \ref{tab:forcing}), for a narrow spherical shell $k_{\rm f} -1 =124 \le k \le k_{\rm f}+1=126$ we add forcing terms obeying the Brownian motion 
$\mathrm{d}F_{\bf k} (t)=
f_0\mathrm{d}{W}_{\bf k}$, where {$W_{\bf k}$} is the Wiener process and in what follows 
$f_0$ is a positive constant. Naturally, $k_{\rm f}$ is taken small for the direct cascade and large for the inverse one.
The dissipation term is of the form $D_{\bf k} = (k/k_{\rm L})^{-2\alpha}+(k/k_{\rm R})^{2\beta}$; it acts at small and/or large scales. Moreover, the condensate mode $k=0$ is also dissipated with constant friction $D_{\bf 0}$ when necessary. 
The WKE is forced by adding to its RHS a constant-in-time function $f_\omega=c_{\rm f}\, \Gamma(\omega)$, where $\Gamma(\omega)$ is the Gaussian centered at $\omega_{\rm f}$ and of width $\Delta\omega_{\rm f}$. Dissipation is introduced by adding the term $-[(\omega/\omega_{\rm L})^{-\alpha}+(\omega/\omega_{\rm R})^{\beta}]n_\omega$ to the RHS of the WKE. 
 We perform two WKE simulations -- one for the direct cascade and another for the inverse one (cases~9 and~10 
 in Table \ref{tab:forcing}, respectively). All the numerical parameters are given in Table \ref{tab:forcing}. 
 \begin{table*}[!hbtp]
\begin{tabularx}{0.98\textwidth}{|X|X|X|X|X|X|X|X|X|X|X|X|X|}
\hhline{|=|=|=|=|=|=|=|=|=|=|=|=|=|}
case&model&cascade&$L$&$N_p$&$n_{\rm f}$ & $f_0^2$ & $k_{\rm f}$ & $D_{\bm{0}}$ &$k_L$& $\alpha$ & $k_R$ & $\beta$ \\
\hline
6 & GPE & direct & $4\pi$ & 512 &    1.5  & --  & 1 &  $10^3$& -- & -- & ---  &  ---  \\ 
7&  GPE & inverse &  $2\pi$  &  512   & -- & $10^{-4}$ &  125 & -- & -- & -- & 130 & 6 \\
8&  GPE &  inverse  &  $2\pi$  &   512  &  --  & $10^{-4}$  & 125  & $10^3$ & 1 & 0.5 & 130  & 6 \\
\hhline{|=|=|=|=|=|=|=|=|=|=|=|=|=|}
case & model & cascade & $\omega_{\rm min}$ & $\omega_{\rm max}$ & $c_{\rm f}$ & $\omega_{\rm f}$ & $\Delta\omega_{\rm f}$ &  $\omega_{\rm L}$ & $\alpha$   &  $\omega_{\rm R}$ & $\beta$ & $k_{\rm f}$\\
\hline
9 & WKE  &direct & $10^{-5}$ & 10 & 10  & $3\times 10^{-4}$ & $3\times 10^{-4}$ & $10^{-4}$ & 4 &  --- & --- & 
$0.025$ \\
10 & WKE & inverse & 0.1 & $10^5$ &  50 & $125^2$ & 500  & 10 & 4 & $10^5/4.5$ & 7  & --- \\
\hline
\end{tabularx}
\caption{\label{tab:forcing}%
Parameters of simulations of forced-dissipated GPE and WKE. 
}
\end{table*}

\subsubsection{Forcing at low $k$'s: direct cascade}

\begin{figure}[hbtp]
\begin{center}
\includegraphics[scale=1]{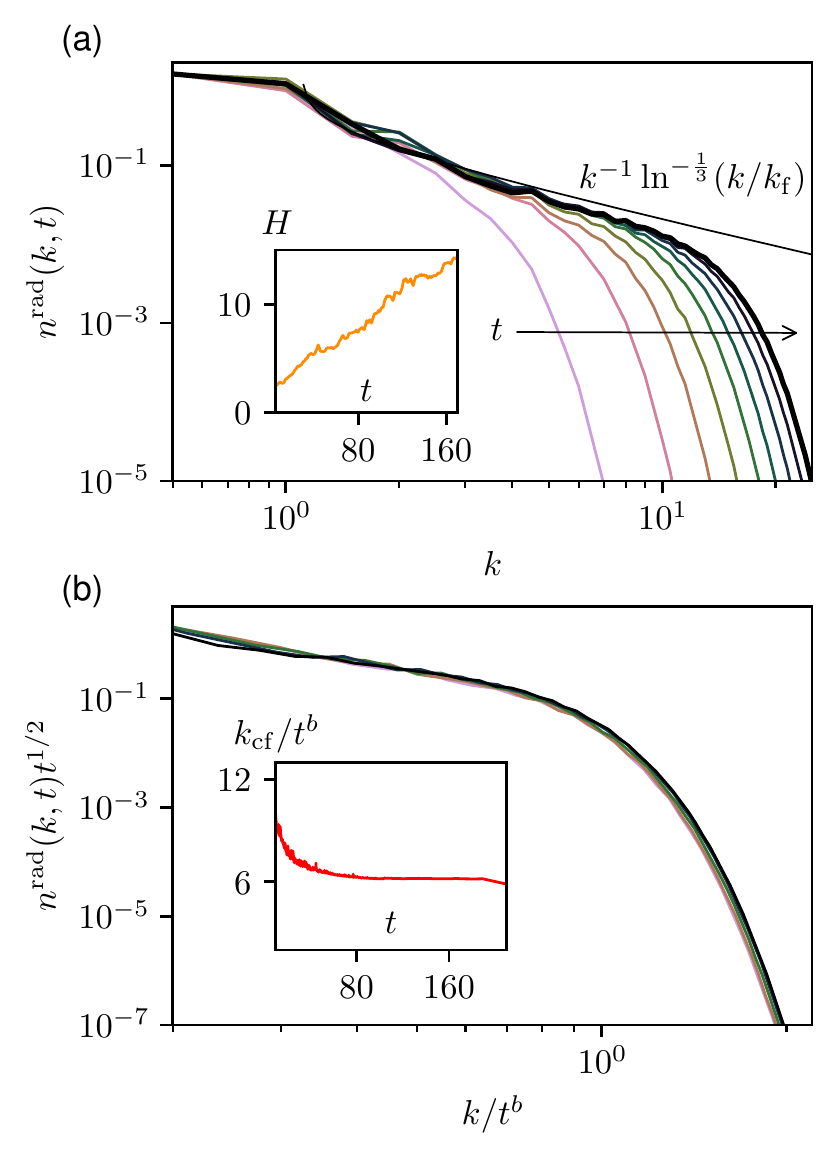}
\caption{Numerical results for GPE simulation (case~6) with forcing at low $k$'s. (a) Values of the spectrum $\nrad (k,t)$ for $t=20\,,50\,,80\,,110\,,140\,,170\,,200\,,230\,, 260$, and in the inset --  time evolution of energy density $H(t)$;   (b) $\nrad (k,t)$ compensated by $t^{1/2}$ vs. compensated wave number $k/t^b$ for $t=80\,, 100\,,120\,,140\,,160$ and in the inset -- time evolution of the wave front $\kcf$ divided by $t^b$. }\label{fig:nk_forc_dirct_gpe}
\end{center}
\end{figure}
\begin{figure}[hbtp]
\begin{center}
\includegraphics[scale=1]{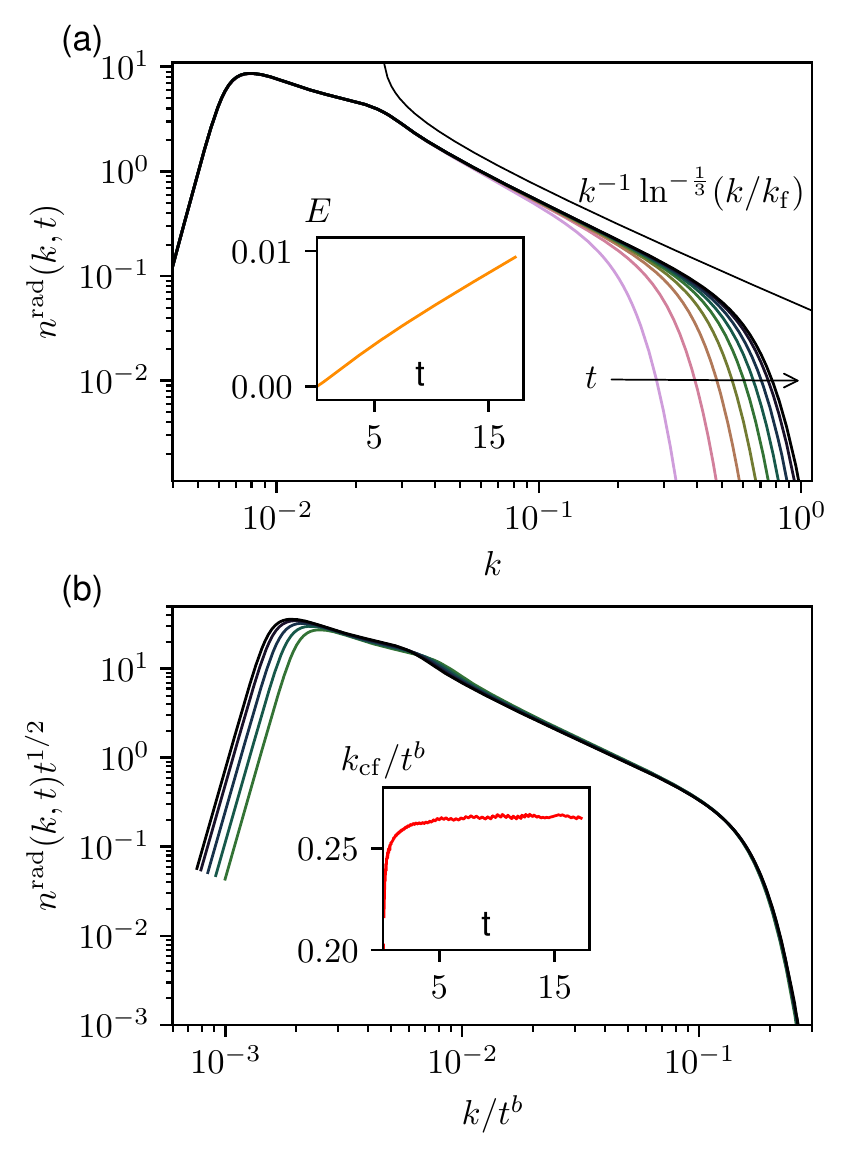}
\caption{Numerical results for WKE simulation (case~9) with forcing at low $k$'s. (a) Values of the spectrum $\nrad (k,t)$ for the times $t=2\,,4\,,6\,,8\,,10\,,12\,,14\,,16\,,17.3$ and in the inset -- time evolution of energy density $E$;   (b) $\nrad (k,t)$ compensated by $t^{1/2}$ vs. compensated wave number $k/t^b$ for the times  $t=10\,,12\,,14\,,16\,,17.3$ and the inset for the time evolution of wave front $\kcf$ divided by $t^b$. }\label{fig:nk_forc_dirct_wke}
\end{center}
\end{figure}

To implement the direct cascade, we simulate the GPE and the WKE  with forcing terms centered at low wave numbers and let the spectrum propagate to large wave numbers (we set no dissipation at large $k$'s).  Figures \ref{fig:nk_forc_dirct_gpe} and \ref{fig:nk_forc_dirct_wke} present numerical results of the GPE and the WKE simulations, respectively (cases~6 and~9 in Table \ref{tab:forcing}).
Observing the time evolution of the wave-action spectrum in Figure~\ref{fig:nk_forc_dirct_wke}~(a), one can see the formation of the log-corrected KZ spectrum behind the front. Thanks to the good scale separation, for this simulation we can also see the quasi-steady state behind the wave number where the forcing is imposed. However, the inverse-cascade range is too short and the  KZ power law $\sim k^{-1/3}$ is not observed in this case.
In Figure \ref{fig:nk_forc_dirct_gpe} (a) for the GPE simulation we also see a curve qualitatively consistent with the log-corrected KZ spectrum which develops behind the front, but the agreement is poorer than for the WKE simulation.
The inset in Figure \ref{fig:nk_forc_dirct_gpe} (a) shows the GPE energy $H(t)$ which is  increasing nearly linear for all the times, whereas the inset in Figure \ref{fig:nk_forc_dirct_wke} (a) shows an increase of the WKE energy $E(t)$ which is almost linear  for $t > 5$. This dependence $E(t)\sim t$
 implies that in the self-similar solution of the first-kind \eqref{eq:self-first} $\lambda=1$. Using the last equality that gives $b=1/2$, we plotted 
the self-similar solutions in Figure \ref{fig:nk_forc_dirct_gpe} (b) for $t\in[100,160]$ and in Figure \ref{fig:nk_forc_dirct_wke} (b) for $t\in[2,17.3]$. In these time intervals, an evident visual collapse of plots can be observed for both GPE and WKE simulations. The time windows, in which we observe clear self-similar evolution, also agree with time windows in which the compensated wavefront $\kcf$ is constant (see the insets). Note that in the GPE simulations, we start to lose self-similarity around $t=180$, because the front touches the right boundary (maximum wave number).

It is worth mentioning that the numerical setup that we used for  the direct cascade, mimics the configuration in the experiment by Navon et al. \cite{navon2019synthetic}, with a constant rate of energy injection and almost steady total number of particles.  
In their study, they reported $b\approx 0.54$ with a tolerance of $6\%$ when converted to our notations. This finding is remarkably close to our theoretical prediction $b=1/2$.

\subsubsection{Forcing at high $k$'s: inverse cascade}

We perform two GPE simulations with the same forcing centered at high $k$'s and hyper-viscosity acting at even higher $k$'s located to the right from the forcing range to achieve the inverse cascade (cases~7 and~8 in Table \ref{tab:forcing}). The only difference  is that for case~8 we put hypo-viscosity at low $k$'s to absorb the inverse cascade of particles and to get a steady state, whereas in case~7 no dissipation at low $k$'s was imposed. Therefore, the two simulations show almost the same behavior before the left parts of the spectral fronts get into the region where the dissipation of case~8 is imposed. 
\begin{figure*}[htbp]
\begin{center}
\centering{\includegraphics[scale=1]{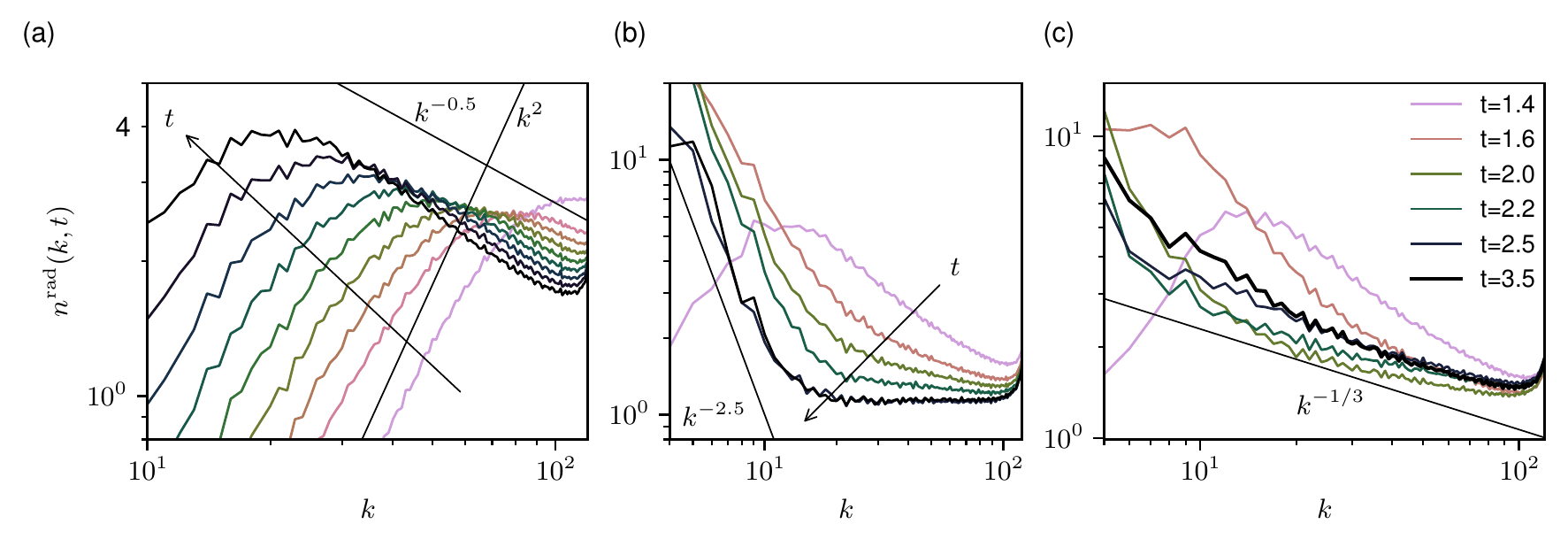}}
\caption{Numerical results for GPE simulations with forcing at high $k$'s: (a) Spectrum $\nrad (k,t)$ (of case~7) for early times $t=0.85\,, 0.95\,, 1\,, 1.05\,, 1.1\,, 1.15\,, 1.2\,,  1.25\,, 1.3$;   (b) Long-time evolution of $\nrad (k,t)$ (of case~7) without dissipation at low $k$ for times $t=1.4\,, 1.6\,, 1.7\,, 1.8\,,  2.4\,,  3$;  (c) Long-time evolution of $\nrad (k,t)$ (of case~8) with dissipation at low $k$'s.}\label{fig:nk_forc_inv_gpe}
\end{center}
\end{figure*}
Figure \ref{fig:nk_forc_inv_gpe} (a) presents the early time  evolution of the wave-action spectra for $t\in[0.85,1.3]$ obtained for case~7. As in the free-decay simulations, we see a particle equipartition scaling $k^2$ in the left front and an anomalous 
$k^{-x^*}$-scaling with $x^*\approx 0.5$
behind the front. The late time evolution of case~7 is plotted in Figure \ref{fig:nk_forc_inv_gpe} (b), where we see the $k^{-2.5}$ scaling for small $k$'s and the energy equipartition for large $k$'s, as it was reported in \cite{shukla2022nonequilibrium}. It is also similar to the late-time evolution of freely decaying inverse cascade plotted in Figure \ref{fig:nk_free_inv_gpe} (a). The final state appears to be quasi-steady with condensate ratio ($\nrad(0,t)/N$) oscillating around $10^{-3}$ -- a picture previously observed in \cite{shukla2022nonequilibrium}. Obviously, for the final state to be steady the particle and energy inputs and sinks in the high-$k$ region must cancel each other.

Let us now consider the long-time evolution in case~8. We expect that due to the presence of the dissipation at low $k$'s the system will eventually relax to the steady KZ spectrum corresponding to the inverse cascade of particles, $\nrad_k \sim k^{-1/3}$. The route to this final steady KZ spectrum is interesting;
it can be seen in the sequence of spectra plotted in Figure \ref{fig:nk_forc_inv_gpe} (c) for different  moments  of the late evolution. Shortly after the anomalous spectrum $\sim k^{-x^*} \approx k^{-0.5}$ 
forms at $t\to t^* \approx 1.5$, we see a spectrum overshoot at low $k$'s before it relaxes to smaller values after  the low-$k$ dissipation takes effect. This is followed by an opposite overshoot characterized by the spectrum depletion and formation of a slope shallower than $k^{-1/3}$. Only after that the spectral 
slope moves up towards and stabilizes at, ${-1/3}$. This oscillatory relaxation to the KZ steady state
is rather different from the reflected wave scenario (corresponding to the self-similarity of the third kind) described at the end of section \ref{Sec:3}.

\begin{figure}[htbp]
\begin{center}
\includegraphics[scale=1]{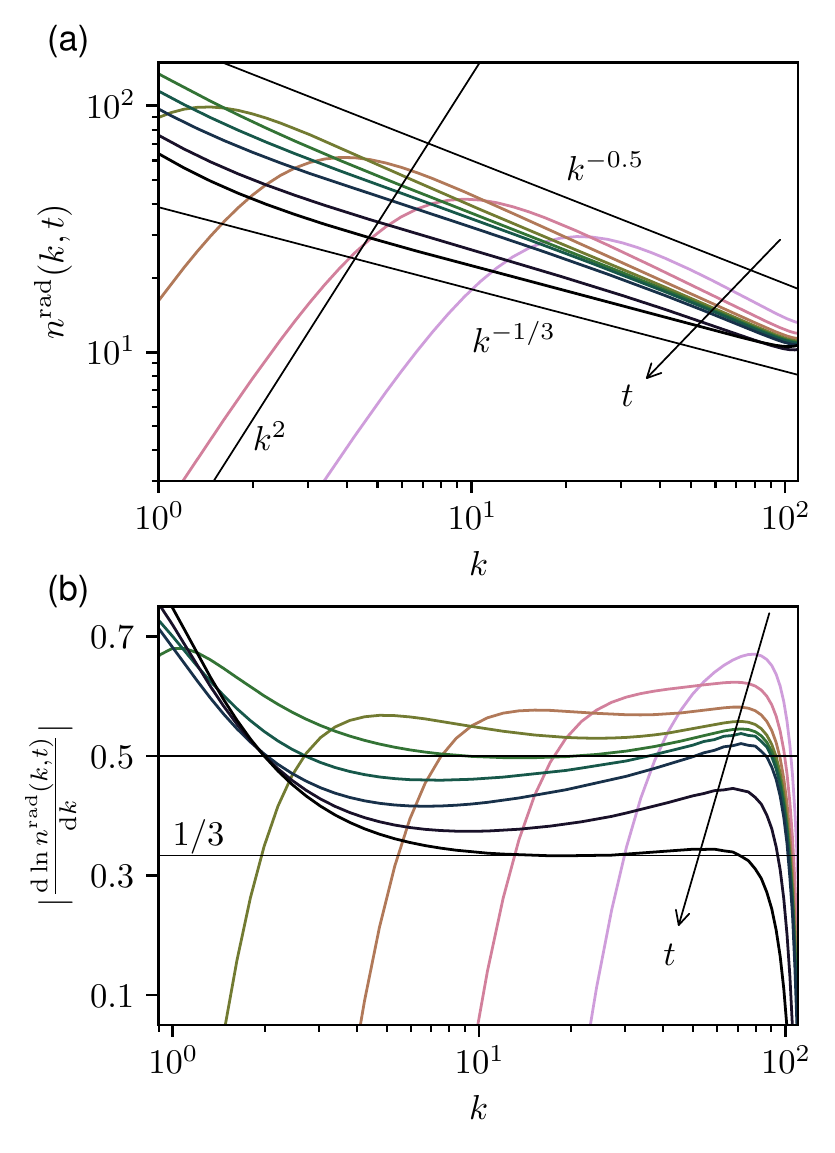}
\caption{ Numerical results obtained by WKE simulation (case~10)
with forcing at intermediate $k$'s and dissipation at both low and high $k$'s: (a) Spectrum $\nrad (k,t)$ for  times $t=0.0176\,, 0.0185\,, 0.0190\,,$ $0.0192\,, 0.0194\,, 0.0195\,, 0.0197\,, 0.0210\,, 0.0379$;   (b) Slope of $\ln{\nradk}$ for the same times as in (a). }\label{fig:nk_forc_inv_wke}
\end{center}
\end{figure}

For the WKE, long-time inverse-cascade evolution exists only if a low-$k$ dissipation is present because otherwise the
spectrum blows up at $t=t^*$. Numerical results for this case (case~10 in Table \ref{tab:forcing}) are presented in Figure~\ref{fig:nk_forc_inv_wke} with
spectra shown in panel (a) and the spectral slopes (log-derivatives of $\nrad_k$) -- in panel (b).
Here, we see that the transition from the slope of anomalous exponent
$-x^*$ ($\approx - 0.5$) to the KZ slope $-1/3$  proceeds monotonously -- faster at lower and slower at higher $k$'s. This is consistent with the reflected wave scenario  described at the end of section \ref{Sec:3}. However, the range of wave numbers achievable in numerics is insufficient for making any conclusions about the realisability of third-kind self-similarity.

\section{Summary and discussion}

In this paper, we studied evolving BEC wave turbulence using numerical simulations of the GPE and the WKE in several different setups corresponding to free-decaying and forced-dissipated cases for developing inverse, direct and dual cascades. Our focus was on identifying self-similar evolution regimes.
In both the GPE and the WKE simulations, we observe the first-kind self-similarity for the direct cascade. In the free-decay simulations, the self-similar spectra tend to stationary thermal energy equipartition states at large times, $t\to \infty$.
The temperature of such final states is determined by the total initial energy in the system. 
For the forced-dissipated setup, the final steady state is the direct-cascade 
KZ spectrum (which forms immediately behind the propagating front of the self-similar spectrum.

For the inverse cascade evolution, we have verified the existence of the self-similar regime of the second kind describing self-accelerating dynamics of the spectrum leading to blowup at $k=0$ at a finite
time $t=t^*$. Due to the fact that close to $t^*$ the nonlinear dynamics is faster than the forcing process,
the self-similar evolution of the second kind is insensitive to the presence or absence of forcing.
Physically, this process describes a non-equilibrium Bose-Einstein condensation or, to be more precise,
pre-condensation -- because at $t=t^*$ the particle occupation of $k=0$ is still zero (i.e. the integral of total number of particles \eqref{eq:cons-Nwke} converges at $k=0$ on the spectrum $k^{-x^*}$).
For WKE, the blowup is exact, meaning that no free-decay evolution can be considered for $t>t^*$
without regularising the WKE model, e.g. via introducing an evolution equation for the $k=0$ mode and coupling it
to the equation for modes with $k > 0$ as was done in \cite{semikoz1995kinetics,semikoz1997condensation}. 
We have not attempted to consider such post-$t^*$ evolution with this kind of regularisation. 
Interestingly, a significant numerical resolution available for the WKE simulations allowed us
to implement a dual cascade free-decay system where both the direct cascade (first-kind self-similarity) and
the inverse cascade (the second-kind self-similarity) are observed simultaneously for times up to $t^*$.
For GPE, the blowup behavior is only an approximate intermediate asymptotic. Closer to $t^*$ the self-similar behavior breaks down both due to the breakdown of the weak nonlinearity assumption and due to the discreteness of the $k$-space corresponding to a finite retaining box. As a result, the GPE evolution  continues regularly past  $t^*$. Namely, in the absence of low-$k$ dissipation, it shows the formation of an energy equipartition spectrum at high $k$'s
and a sharp spectrum in the low-$k$ region -- a condensate. We emphasize the fact that in the case
of the GPE system  in a finite (periodic) box, the condensate is spread over few lowest-$k$ modes,
and not concentrated at $k=0$ only as in the WKE case. In the presence of low-$k$ dissipation, in both 
the WKE and the GPE systems,
the condensate is suppressed and the spectrum relaxes to the KZ inverse cascade steady state. For the GPE,
an oscillation is observed in the transient period, whereas for the WKE system (which is now regularised by the low-$k$ dissipation and can evolve for $t>t^*$)  we see signatures of the reflected wave scenario characterized by the third-type self-similarity.
However, these signatures are quite indirect and more work is needed for identifying the 
third-type self-similarity which, in our case, still remains hypothetical.
Besides numerical simulations for a much wider range of $k$, which would presently be hard to achieve,
one could solve directly the integro-differential equation for the self-similar shape in a way
similar to the one used for a three-wave kinetic equation for MHD waves in \cite{Bell_2018}.
This could be an interesting problem for future study.

Finding universal dynamics in turbulent superfluid Bose gases has gained significant interest in recent laboratory experiments (see \cite{navon2019synthetic, glidden2021bidirectional, garcia2022universal, madeira2023differential}). These experiments involve the self-similar regimes discussed in the paper. 
In particular, the experiments detailed in \cite{navon2019synthetic} and \cite{glidden2021bidirectional} have reported instances of the first kind of self-similarities for the direct cascade. These cases encompass scenarios involving free-decay with nearly constant energy and forced cases with linearly increasing energy, respectively. Remarkably, the
exponent constants observed in these experiments closely align with our predictions by Equation \eqref{eq:self-first} for both configurations.
However, the experimental investigation of the inverse cascade poses a more intricate challenge than the direct cascade. Recent experiments conducted in \cite{glidden2021bidirectional}, \cite{garcia2022universal}, and \cite{madeira2023differential} aimed to detect self-similarity of the first kind. It is important to note that, as clarified in the current paper, in the inverse cascade, the second kind of self-similarity should be expected instead, and therefore, the respective experiments should be revised and refined.

Finally, special interest for future studies presents 2D setups relevant to optical turbulence systems, see, e.g., \cite{Baudin2020Condensation}. 
Note that 2D systems are special and different from 3D ones since neither true condensation nor pure KZ cascades are possible for such systems.
Also, since the 2D optical systems are usually trapped by a carrier beam, the system may remain weakly nonlinear even if a significant part of the total particles accumulate in the ground energy state, as it was the case in experiment of
\cite{Baudin2020Condensation}.
Evolving 2D BEC WT is also very different from the 3D evolving turbulence, requiring a separate theoretical investigation.

\section{Acknowledgements}
This work was funded by the Simons Foundation Collaboration grant Wave Turbulence (Award ID 651471). 
This work was granted access to the high-performance computing facilities under GENCI (Grand Equipement National
de Calcul Intensif) A0102A12494 (IDRIS and CINES), the
OPAL infrastructure from Université Côte d’Azur, supported
by the French government, through the UCAJEDI Investments in the Future project managed by the National Research
Agency (ANR) under Reference No. ANR-15-IDEX-01. The work of Boris Semisalov was supported by the RSF (Agreement No.~22-11-00287)

\appendix
\section{Naive truncation of the wave kinetic equation \label{App:Naivetruncation}}

\begin{figure}[htbp]
\begin{center}
\centering{\includegraphics[scale=1]{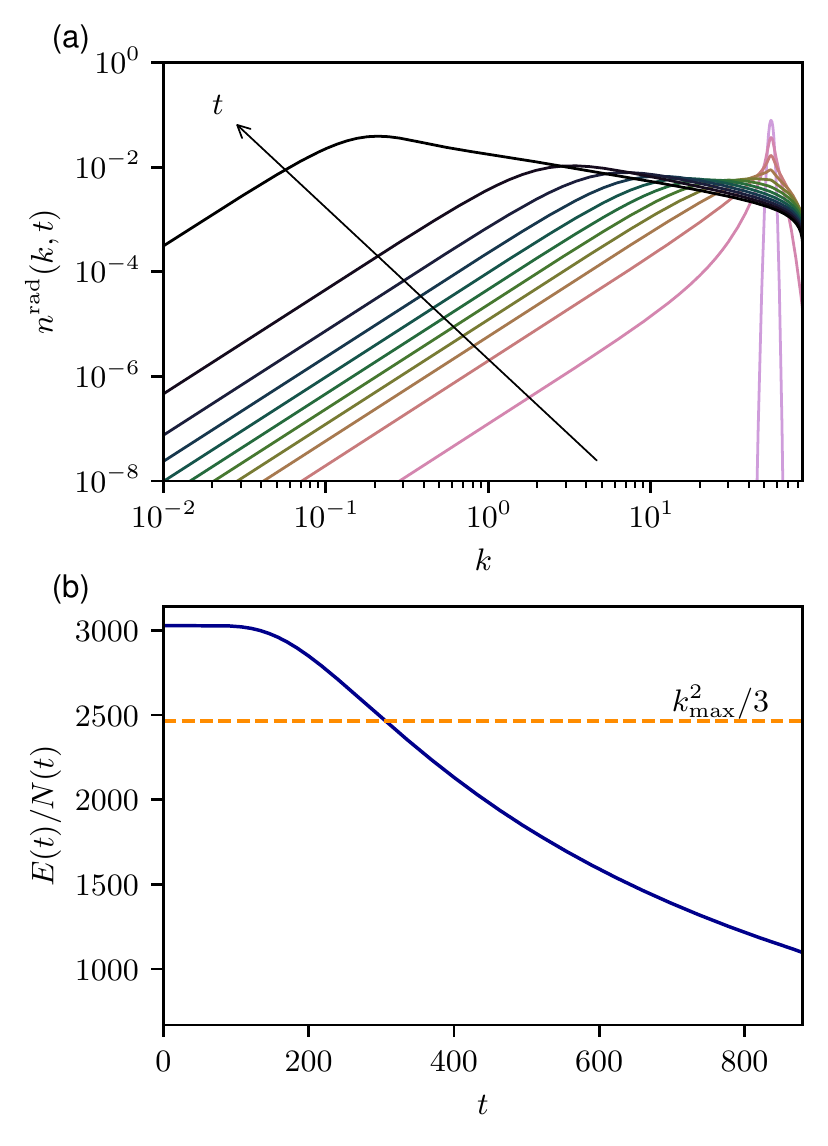}}
\caption{Numerical results for WKE with the leak of energy: (a) Spectrum $\nrad (k,t)$ for  times $t=0, 80, 160, 240,...,880$;   (b) Relation $E/N$ depending on time.}\label{WKE_BU}
\end{center}
\end{figure}

As discussed in section \ref{subsubsec:truncatedWKE}, a proper truncation is needed in order to conserve the invariants. In this Appendix, we show the spurious consequences of the naive truncation. Namely, we impose the cutoff $\omega_{\rm max}$ in the collision term, but keep the subdomains $\Omega_2$, $\Delta_1$ and $\Delta_2$ sketched in Figure~\ref{IntDomain}.

We repeat the simulation of case~2 in Table \ref{tab:free} using this truncation scheme. Despite the facts that for such conditions $E/N \approx3028 > k^2_{max}/3\approx2465$ and the initial spectrum does not touch $\omega_{\rm max}$, we got blowup in a finite time, see Figure~\Ref{WKE_BU}(a).

The reason for this effect can be extracted from Figure~\Ref{WKE_BU}(b). The relation $E/N$ starts to decrease rapidly at $t\sim 150$, and for $t = 300$ we already have $E/N < k^2_{max}/3$. Let $t_c$ be the time when the equality $E/N = k^2_{max}/3$ is reached. In Figure~\Ref{WKE_BU}(a) one can see that for $t<t_c$ the evolution of spectrum slows down and RJ spectrum starts to develop, but for $t>t_c$ the evolution accelerates and spectrum goes to blowup. The reason is that the condition $E/N < k^2_{max}/3$ is dynamically broken as a consequence of the energy leak.

This spurious evolution is  an effect of the naive UV-cutoff only and, by using finer frequency and time discretizations, we have verified that  it is not caused by finite grid effects or time stepping.

\bibliographystyle{apsrev4-2}
\bibliography{apssamp}

\end{document}